\documentstyle[epsfig,twoside,rotating]{aa}

\begin{document}

\title{Does transparent hidden matter
generate optical scintillation?}
\author{
M.~Moniez\inst{1}
}
\institute{
Laboratoire de l'Acc\'{e}l\'{e}rateur Lin\'{e}aire,
{\sc IN2P3-CNRS}, Universit\'e de Paris-Sud, B.P. 34, 91898 Orsay Cedex, France
}

\offprints{M. Moniez, \email{ moniez@lal.in2p3.fr}}

\date{Received 17 February 2003 / Accepted 8 September 2003}
%
%

\abstract{
Stars twinkle because their light goes through the atmosphere.
The same phenomenon is expected when the light of extra-galactic stars
goes through a Galactic -- disk or halo -- refractive medium.
Because of the large distances involved here, the length and time
scales of the optical intensity fluctuations resulting from the
wave distortions are accessible to the current technology.
In this paper, we discuss the different possible scintillation
regimes and we focus on the so-called strong diffractive regime
that is likely to produce large intensity contrasts.
The critical relationship between the source angular size and the
intensity contrast in optical wavelengths is also
discussed in detail.
We propose to monitor small extra-galactic stars every $\sim 10\,\mathrm{s}$
to search for intensity scintillation produced by molecular hydrogen
clouds.
We discuss means to discriminate such hidden matter
signal from the foreground effects on light propagation.
Appropriate observation of the scintillation process
described here should allow one to detect column density
stochastic variations in Galactic molecular clouds of order of
$\sim 3\times 10^{-5}\,\mathrm{g/cm^2}$, that is
$10^{19}\,\mathrm{molecules/cm^2}$
per $\sim 10\,000\,\mathrm{km}$ transverse distance.
      
\keywords{Cosmology: dark matter -- Galaxy: disk -- Galaxy: halo -- ISM: clouds -- ISM: molecules -- Diffusion}
}

\maketitle

\section{Introduction: The Galactic hidden matter problem and the baryons}
The study of rotation curves of spiral galaxies has
led to the hypothesis of
massive extended halos of ``dark matter" (e.g. \cite{primac}).
From the big-bang nucleosynthesis theory and the measured primordial
abundances of the light elements, it is
also established that baryons in the Universe are at least ten times
more abundant than in the visible matter
(stars, dust and gas)(\cite{Olive}).
This deficit of baryonic matter in the Universe is approximately equal
to the deficit of gravitational matter in our Galaxy. This coincidence
was one of the major motivations for the first microlensing searches
for baryonic hidden matter under the form of Massive Compact Halo Objects
(MACHOs).
Considering the results of these searches
(\cite{notenoughmachos}; \cite{SMC5ans}; \cite{MACHO}),
it seems that the only constituent
that could contribute quite significantly
to the Galactic baryonic hidden matter is the
cool molecular hydrogen ($\mathrm{H_2}$).
Indeed, this material is very difficult to detect, due to the
symmetry of the $\mathrm{H_2}$ molecule that cancels the dipolar
electric transitions, the only ones to be excited in a cold medium
($\sim 10\,\mathrm{K}$).
The perspectives for detecting cold $\mathrm{H_2}$ in outer
Galactic disks have been reviewed by \cite{perspectives}.
Most of the techniques currently used to estimate $\mathrm{H_2}$ Galactic contribution
are indirect detections using tracers such as CO molecule or dust, that
imply specific hypotheses. A possible
direct method using the ultra-fine transitions
of the nuclear spins (from parallel to anti-parallel)
concerns radio emission or absorption ($\lambda=0.5\,\mathrm{km}$ and $5.5\,\mathrm{km}$)
that is impossible to detect in the terrestrial environment.
Detection of Galactic
$\mathrm{H_2}$ clouds in front of quasars is not easily feasible, because
the absorption lines have wavelengths shorter than $110\,\mathrm{nm}$.
As a summary, the hypothesis
of a leading contribution to the halo mass due to cold molecular hydrogen
is not yet ruled out by any of the currently used methods.
A hierarchical structure for cold molecular hydrogen
has been suggested by \cite{fractal} to fill the Galactic thick disk
and \cite{Jetzer1} (1995 and 1998) have considered cold molecular
hydrogen clouds as possible candidates for the Galactic halo dark matter.
According to \cite{fractal} model,
the gas could form ``clumpuscules'' of $10\,\mathrm{AU}$ size at the smallest scale,
with a column density of $10^{24-25}\,\mathrm{cm^{-2}}$, and a surface filling factor
less than 1\%.
We propose to search for such cool molecular clouds
in the thick disk and in the halo through their
diffraction and refraction effects on the propagation of remote stars light.
The method that will be discussed in this paper is of more general
use and could be used to
detect other types of transparent structures. Nevertheless,
our immediate aim is to demonstrate its feasibility and its sensitivity
to gaseous structures that are considered as viable candidates for the
Galactic hidden matter.
\section{Detection mode of extended $\mathrm{H_2}$ clouds: principle}
Due to index refraction effects,
an inhomogeneous transparent (gaseous) medium
distorts the wave-fronts of incident
electro-magnetic waves.
When observing a remote source located behind a gaseous structure
(hereafter called screen), the luminous amplitude results from
the propagation of the distorted wave-front.
Diffraction theory predicts that
interference patterns should form and also possibly
longer scale refraction effects (prism and lens-like).
In the following, we will study the visibility of the wave-front
distortions
through the intensity contrasts produced on Earth.
We will particularly discuss the critical aspect of the
spatial coherence and show that only the light of
faint stars (remote and small) can provide a detectable
signal of intensity variations due to a refractive structure.

Before developing the scintillation mechanisms, we
stress that the
effect of a refraction process is cumulative, in the sense that
the more material the light encounters, the larger are the
wave distortions.
If there is no dust, the sources behind a gaseous structure are always
visible provided that the wavelength is not resonant with a transition
process of the medium constituents.
\section{Refraction by Galactic hidden gas}
The elementary process responsible for the refraction index
effect is the polarizability of molecules.
After propagation along a distance $l$,
the optical path difference between the vacuum and a medium
characterized by a number density $N$
(number of molecules per volume unit)
and polarizability $\alpha$ is
\begin{equation}
\delta=2\pi Nl\alpha.
\end{equation}
It depends only on the total column number density $Nl$ of
molecules and on their {\it average} polarizability $\alpha$,
but not on the
details of the density distribution along the propagation path.
In the present case, the density is so low
($\sim 10^9\,\mathrm{molecules/cm^3}$)
that the medium cannot be considered as continuous at the
optical wavelength scale (the average distance between molecules
is $\sim 10\,\mathrm{\mu}\gg \lambda$).
Then it is not possible to consider that the polarizability
in a small region (compared with the optical wavelength) results
from the average of polarizabilities over the molecule orientations
as it is usually considered in dense mediums\footnote{As the
$\mathrm{H_2}$ molecule is not spherical,
its polarization may not be aligned with the electric field, and the
polarization averaged over the molecule orientations is
different than the individual polarizations.}.
Nevertheless, we study the result of the diffusion at extremely
large distances from the screen and precisely
towards the initial propagation direction,
where the diffusion is coherent;
then the collective effect of successive diffusions by randomly
oriented molecules
scattered along the propagating path is the same than the diffusion by
molecules with the average polarizability (\cite{hamilton}).

For a gas only made of $\mathrm{H_2}$ molecules,
$\alpha=0.802\times 10^{-24}\,\mathrm{cm^3}$ (\cite{Handbook})\footnote{For
a $\mathrm{H_2/He}$ mixing with $24\%$
He by mass -- corresponding to the primordial abundances --
$<\alpha>=0.720\times 10^{-24}\,\mathrm{cm^3}$, and
subsequent calculations should be renormalized accordingly.}.

If the Galactic halo is as described by \cite{caldwell}
(the so-called standard spherical halo) and if it is
completely made of $\mathrm{H_2}$, then the average column density from the
Sun to LMC (respectively SMC and M31) is $0.0253\,\mathrm{g/cm^2}$
(resp. 0.0307 and 0.0206), 
corresponding to $Nl=7.62\times 10^{21}$ molecules per $\,\mathrm{cm^2}$
(resp. $9.25\times 10^{21}$ and $6.20\times 10^{21}$)
or to a column of $2.83\,\mathrm{m}$ (resp. $3.44\,\mathrm{m}$ and $2.31\,\mathrm{m}$)
of $\mathrm{H_2}$ under normal pressure and temperature conditions.
The extra optical path induced by this medium with respect to the
vacuum is $\delta\simeq 0.4\,\mathrm{mm}\simeq 768\lambda$ at
$\lambda=500\,\mathrm{nm}$ towards LMC (resp. $932\lambda$ and $625\lambda$).
The same orders of magnitude are expected
if hidden matter lies in the thick disk instead of the halo.

These values are average values and we have to take into account
the structuration of this gas.
In the Pfenniger-Combes model, the smallest $\mathrm{H_2}$
structures are $10\,\mathrm{AU}$ wide and have a Jupiter mass.
Therefore their surface filling factor is less than 1\%.
This means that we expect concentration factors of
the column density of 100 at least ($\Rightarrow Nl\sim 10^{24}\,\mathrm{cm^{-2}}$)
for 1\% of the sky fields.
For such a structure
the {\it average} transverse gradient of optical path differences is of
order of
$$\frac{800\times \lambda\times 100\ (concentration)}{5\,\mathrm{AU}}\sim 1\lambda\ per\ 10\,000\,\mathrm{km},$$
for $\lambda=500\,\mathrm{nm}$.
As we will see in the next section,
this fits with the typical optical path fluctuations that
can produce interference patterns
for a $\mathrm{H_2}$ cloud located in the Galactic disks or Galactic halo.
\section{Diffractive scintillation:
Fresnel diffraction applied to a stellar source and a Galactic
screen}
\subsection{General formalism, notations}
\begin{figure}
\begin{center}
\begin{turn}{0}
\mbox{\epsfig{file=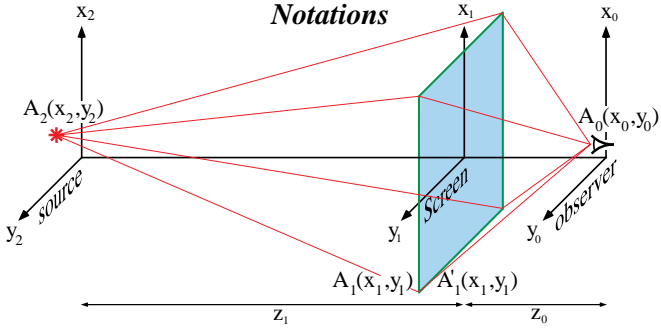,width=8.7cm}}
\end{turn}
\caption[]{Notations: The source is located in the $(x_2,y_2)$ plane,
the screen contains
the diffusive structure, and the observer is located
in the $(x_0,y_0)$ plane.
$A_1(x_1,y_1)$ and $A'_1(x_1,y_1)$ are the amplitudes before and
after screen crossing.
\label{notations}}
\end{center}
\end{figure}
Let $A_2(x_2,y_2)$ be the luminous (complex) amplitude produced
in the source plane (see Fig. \ref{notations}).
For a monochromatic (wavelength $\lambda$)
point-like source of amplitude $A_2$
located at $(x_2,y_2)$ in the source plane,
omitting the time periodic factor $e^{-i\omega t}$,
the amplitude on the screen before diffusion is given by the
spherical wave equation:
\begin{equation}
A_1(x_1,y_1)=A_2 e^{ikr_{12}}/r_{12}\simeq A_2 e^{ikr_{12}}/z_1,
\end{equation}
where $\bf{k}$ is the wave vector ($k=2\pi/\lambda$), and
\begin{eqnarray}
\label{r12}
r_{12}	&=& \sqrt{z_1^2+(x_1-x_2)^2+(y_1-y_2)^2}\\
	&=&z_1\sqrt{1+(\frac{x_1-x_2}{z_1})^2+(\frac{y_1-y_2}{z_1})^2} \nonumber
\end{eqnarray}
can be approximated
by $z_1$ in the denominator because we will always be in the situation
where $z_1\gg x_1$ and $y_1$.
The effect of the screen on the wave propagation can be represented
by a phase delay that depends only on $x_1$ and $y_1$:
\begin{equation}
A'_1(x_1,y_1)=\frac{A_2 e^{ikr_{12}}}{z_1} e^{ik\delta(x_1,y_1)},
\end{equation}
where $\delta(x_1,y_1)$ is the extra optical path due to the screen.
The amplitude after subsequent propagation in vacuum is given by the
Huygens-Fresnel diffraction principle:
\begin{equation}
A_0(x_0,y_0)=\int_{-\infty}^{+\infty}\int_{-\infty}^{+\infty}A'_1(x_1,y_1)
\frac{e^{ikr_{01}}}{i\lambda z_0}dx_1dy_1\ ,
\label{huygens}
\end{equation}
where
\begin{equation}
\label{r01}
r_{01} = z_0\sqrt{1+(\frac{x_0-x_1}{z_0})^2+(\frac{y_0-y_1}{z_0})^2}.
\end{equation}
\subsection{Approximations}
The stationary phase approximation states that the main contribution
to integral (\ref{huygens}) comes from the $(x_1,y_1)$ domain where the phase
term of the integrand does not oscillate too fast.
In this domain, the Fresnel approximation is valid.
It consists in keeping only the first
order development
of the square root in the expression of $r_{01}$:
\begin{equation}
\label{fresnel}
r_{01}\simeq
z_0\left[1+\frac{1}{2}(\frac{x_0-x_1}{z_0})^2+
\frac{1}{2}(\frac{y_0-y_1}{z_0})^2\right].
\end{equation}
This approximation is in principle only valid if the next order terms
are negligible. But as the integrand oscillates very fast
as soon as $(x_0-x_1)^2+(y_0-y_1)^2>\lambda z_0$, the contribution
of the integral vanishes before the Fresnel approximation fails. This is why
we can keep the infinite limits of the integration domain, without the
``artificial'' need of a pupil\footnote{This property has been checked
in the simple example considered below.}.
Within the framework of Fresnel approximation, the
amplitude in the observer's plane is
\begin{eqnarray}
\lefteqn{A_0(x_0,y_0)=} \label{general}\\
& &\frac{e^{ikz_0}}{i\lambda z_0}
\int\!\!\int_{-\infty}^{+\infty}A'_1(x_1,y_1)
e^{\frac{ik}{2 z_0}[(x_0-x_1)^2+(y_0-y_1)^2]}dx_1dy_1 \nonumber \\
&=& \frac{e^{ikz_0}}{2i\pi R_F^2}
\int\!\!\int_{-\infty}^{+\infty}A'_1(x_1,y_1)
e^{i\frac{(x_0-x_1)^2+(y_0-y_1)^2}{2 R_F^2}}dx_1dy_1\ , \nonumber
\end{eqnarray}
where $R_F=\sqrt{z_0/k}=\sqrt{\lambda z_0/2\pi}$ is the Fresnel radius.
$R_F$ is of order of $1500\,\mathrm{km}$ to $15\,000\,\mathrm{km}$ at $\lambda=500\,\mathrm{nm}$,
for a screen located between $1\,\mathrm{kpc}$ to $100\,\mathrm{kpc}$.
This length scale characterizes the $(x_1,y_1)$ domain that contributes
to the integral (a few Fresnel radii).
Then it is clear that the Fraunhofer approximation -- that assumes the
contribution of $(x_1^2+y_1^2)/2 R_F^2$
to the phase term development as a constant over the pupil --
is not valid here.
In return, as we will consider sources located much further
from the observer than the screen,
we can approximate the spherical incident wave to a plane wave.
For a point-like source located at $(x_2,y_2)=(0,0)$,
we therefore assume that
$A_1(x_1,y_1)=A$ is a real constant just before the screen.
The amplitude in the observation plane is then
\begin{eqnarray}
\label{amplit}
\lefteqn{A_0(x_0,y_0)=} \\
& & \frac{Ae^{ikz_0}}{2i\pi R_F^2}
\int\!\!\int_{-\infty}^{+\infty}
e^{ik\delta(x_1,y_1)}
e^{i\frac{(x_0-x_1)^2+(y_0-y_1)^2}{2 R_F^2}}dx_1dy_1\ . \nonumber
\end{eqnarray}
The resulting amplitude is affected by strong interferences
in the observer's plane (the so-called speckle) if 
$\delta(x_1,y_1)$ varies stochastically of order of $\lambda$
within the Fresnel domain.
This is precisely the same order of magnitude as
the average gradient that characterizes the cool $\mathrm{H_2}$ structures.

The above expression corresponds to a point-like monochromatic source.
We will now consider a simple configuration to discuss
spatial and time coherence effects
that severely limit the visibility of diffraction fringes.
\section{Diffractive scintillation:  a simple, pedagogical
(and realistic) screen example}
\label{simplecase}
Fresnel formalism cannot provide a simple procedure to
take into account the source size. In this section, we
analyze the effect of the source size in the
case of a simple phase screen.
Let us assume that the screen is a step of optical path $\delta$
parallel to the $y_1$ axis,
described by a Heaviside distribution in the $(x_1,y_1)$ plane
$\delta(x_1,y_1)=\delta \times H(x_1)$.
This case is realistic since, at the Fresnel scale, the edge of
a gaseous structure can be considered as a straight line that divides
the plane into two regions. The step approaches the effect of a
``strong'' local gradient of the optical path.
The case of a ramp instead of a step has also been examined, but
we will only show here an example of diffraction pattern, because
the calculation of the source size effect is more complicated
in this case.
A more complete discussion is proposed below, that concerns a
wide domain of screen models with stochastic optical
path variations. Nevertheless, this example will be our guide
for feasibility studies.
\subsection{Point-like, monochromatic source}
\begin{figure}
\begin{center}
\begin{turn}{0}
\vbox{\epsfig{file=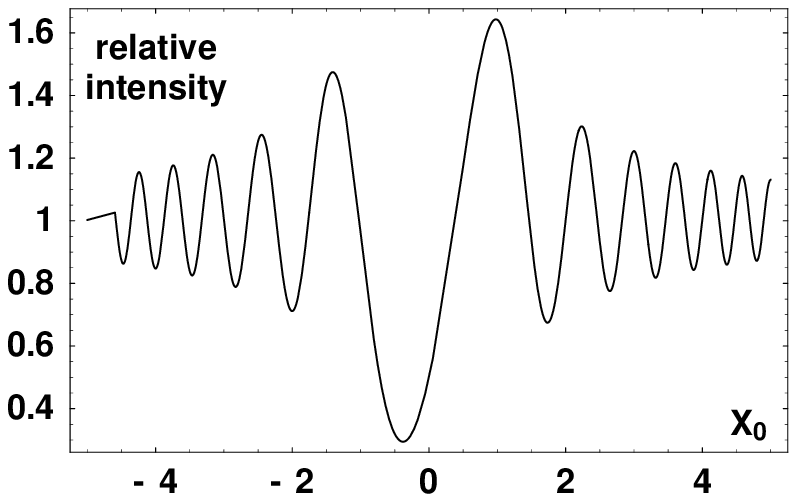,width=8.0cm}
\epsfig{file=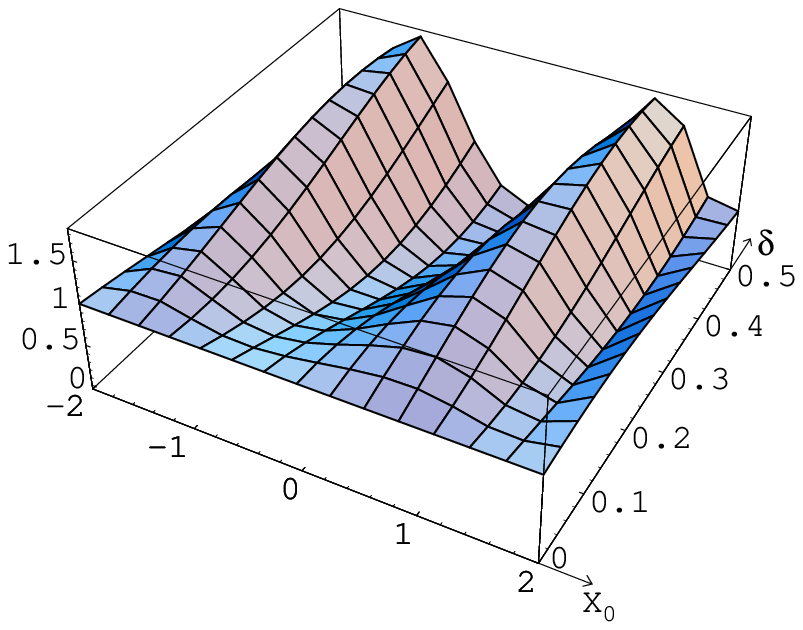,width=8.0cm}
\epsfig{file=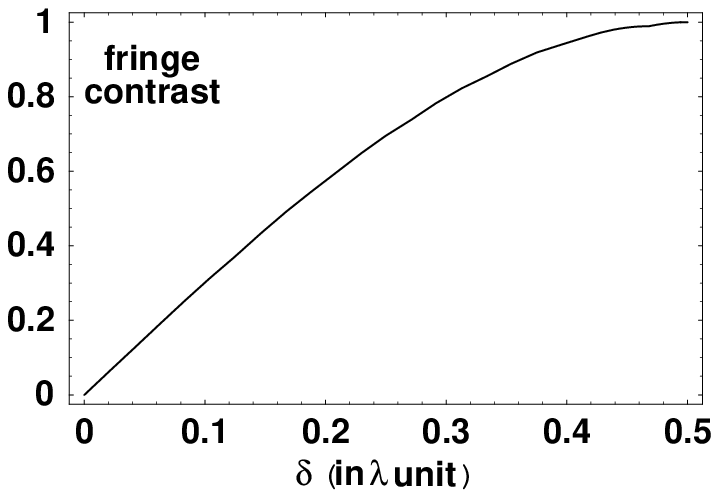,width=8.0cm}}
\end{turn}
\caption[]{{\it Top}: Diffraction pattern produced in the observer's plane,
perpendicularly to a step of optical path $\delta=\lambda/4$.
The $X_0-\mathrm{axis}$ origin is the intercept of the source-step
line with the observer's plane.
$X_0=1$ corresponds to $x_0=\sqrt{\pi}R_F$.\\
{\it Middle}: Diffraction pattern as a function of the step size
$\delta$ (in depth, from $0.$ to $0.5$ in $\lambda$ units).\\
{\it Bottom}: Contrast of the diffraction pattern
as a function of the step size $\delta$.
\label{diffpoint}}
\end{center}
\end{figure}

\begin{figure}
\begin{center}
\begin{turn}{0}
\vbox{\epsfig{file=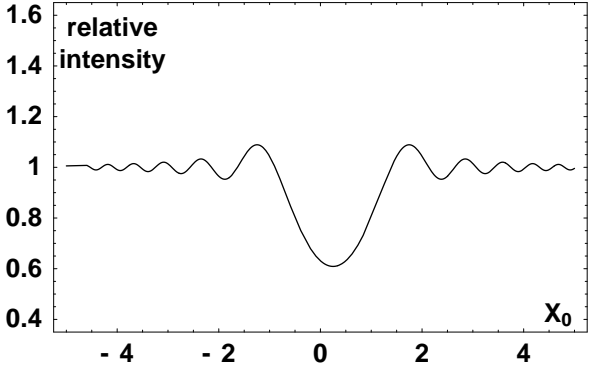,width=8.0cm}
}
\end{turn}
\caption[]{Diffraction pattern
produced perpendicularly to a prism of optical path
with $\delta=X_0\times\lambda/2$ for $X_0>0$.
The $X_0-$axis origin is the intercept of the source-corner
line with the observer's plane.
$X_0=1$ corresponds to $x_0=\sqrt{\pi}R_F$.
\label{diffcoin}}
\end{center}
\end{figure}

The integral (\ref{amplit}) that corresponds to a monochromatic
point-source can easily be separated into a product
of two integrals.
The integral along $y$ can be estimated by noticing that 
in absence of screen there should be no effect on the propagation.
We thus get the following relation
\begin{equation}
\left[\int_{-\infty}^{+\infty}
e^{i\frac{(x_0-x_1)^2}{2 R_F^2}}dx_1\right]^2=2i\pi R_F^2\ .
\end{equation}
Splitting the integral along $x$ into two parts (for $x_1<0$ and $x_1>0$)
we get the expression
\begin{eqnarray}
\lefteqn{A_0(x_0,y_0)=} \\
& & Ae^{ikz_0}+\frac{Ae^{ikz_0}}{\sqrt{2i\pi} R_F}\int_{0}^{+\infty}
\left[e^{ik\delta}-1\right]e^{i\frac{(x_0-x_1)^2}{2 R_F^2}}dx_1\ . \nonumber
\end{eqnarray}
It follows that the amplitude in the observer's plane can be expressed by:
\begin{eqnarray}
\lefteqn{A_0(x_0,y_0)= Ae^{ikz_0}\times} \\
& \left[1+\frac{e^{ik\delta}-1}{2}
[1+S(X_0)+C(X_0)-i(S(X_0)-C(X_0))]\right] \nonumber,
\end{eqnarray}
where $X_0$ is the reduced variable, defined by
\begin{equation}
X_0=x_0/(\sqrt{\pi}R_F),
\end{equation}
S and C are the Fresnel integrals
\begin{eqnarray}
S(X)=\int_{0}^{X}\sin\frac{\pi t^2}{2}dt,\\
C(X)=\int_{0}^{X}\cos\frac{\pi t^2}{2}dt.
\end{eqnarray}
The intensity in the observer's plane is
\begin{equation}
I_0(x_0,y_0)=A_0(x_0,y_0)\times A_0^*(x_0,y_0)=A^2\times i_0(X_0)\ ,
\end{equation}
where
\begin{eqnarray}
\label{intens}
i_0(X_0)&=
& 1-(S(X_0)-C(X_0))\sin(k\delta)+ \\
& & \left[S(X_0)^2+C(X_0)^2-\frac{1}{2}\right][1-\cos(k\delta)]\ . \nonumber
\end{eqnarray}
Fig. \ref{diffpoint} displays the variations of this intensity in
the observer's plane for $\delta=\lambda/4$, and shows
the contrast of the diffraction pattern as a function of the
step size $\delta$. The inter-fringe is --in a natural way--
close to the length scale defined by $\sqrt{\pi}R_F$.
Fig. \ref{diffcoin} also shows the diffraction pattern produced
by a prism whose edge is the $y_1$ axis,
with an optical path variation
$\delta=\lambda/2$ per one unit of $\sqrt{\pi}R_F$.
Such a gradient is comparable with the average gradient
expected from the width variation of $10\,\mathrm{AU}$ gas structures
in the Galactic thick disk.
One should notice here that diffraction patterns
take place if the second derivative of the optical path
is different from zero. Discontinuity (of the optical path or of its
derivatives, as in our examples) is not necessary to get such patterns.
\subsection{Disc-source. Spatial coherence}
\label{spatialcoh}
A simple geometrical construction suggests that
the diffraction pattern produced in the observer's plane
by a point source
located at $(x_2,y_2)$ is the same as the one produced
by a point source at the origin, but translated by
$(-x_2 z_0/z_1,-y_2 z_0/z_1)$.
Actually, it is easy to check that, {\it within the Fresnel approximation}
-- using expressions (\ref{r12}), (\ref{r01}), (\ref{fresnel}) -- ,
the difference between the phase $k r_{01}+k r_{12}+\delta(x_1,y_1)$
along the optical path defined by $(x_2,y_2)$, $(x_1,y_1)$ and $(x_0,y_0)$
and the phase along the path defined by
$(0,0)$, $(x_1,y_1)$ and $(x_0+x_2 z_0/z_1,y_0+y_2 z_0/z_1)$ is
independent of the point $(x_1,y_1)$.
This constant phase difference term can be factorized in
integral (\ref{general}),
and it follows that it is the only difference between
the amplitude
diffracted at $(x_0,y_0)$ from a source located at $(x_2,y_2)$
and the one diffracted at
$(x_0+x_2 z_0/z_1,y_0+y_2 z_0/z_1)$ from a source at the origin.
The intensity of the
diffracted wave from a point-source located at $(x_2,y_2)$ is then given by
\begin{eqnarray}
& I_0(x_0,y_0)[source\ at\ (x_2,y_2)]= \\
& I_0(x_0+x_2\frac{z_0}{z_1},y_0+x_2\frac{z_0}{z_1})[source\ at\ (0,0)]. \nonumber
\end{eqnarray}
An astrophysical source can (almost always) be considered as a
uniform disk (radius $r_s$) of incoherent point-source elements.
We then have
to integrate the intensities from the source elements to get the
resulting diffraction pattern.
As elements aligned parallel to the y-axis give the
same diffraction pattern, the total intensity is given by
\begin{equation}
{\cal I}(x_0,y_0)=\frac{I}{\pi r_s^2}\int_{-r_s}^{r_s}
2\sqrt{r_s^2-x_2^2}\times i_0\left(\frac{x_0+x_2\frac{z_0}{z_1}}{\sqrt{\pi}R_F}\right)dx_2
\label{etendu}
\end{equation}
where $I$ is the total intensity emitted by the object on the screen and
$i_0$ is given by expression (\ref{intens}).
Using the reduced variables
$$X_0=\frac{x_0}{\sqrt{\pi}R_F}\ \ {\rm and}\ \ X_2=\frac{x_2}{\sqrt{\pi}R_F}\times \frac{z_0}{z_1}$$
leads to
\begin{equation}
{\cal I}(x_0,y_0)=\frac{2I}{\pi R_S}\int_{-R_S}^{R_S}
\sqrt{1-\left[\frac{X_2}{R_S}\right]^2} i_0(X_0+X_2)dX_2\ ,
\end{equation}
where
\begin{equation}
R_S=\frac{r_s}{\sqrt{\pi}R_F}\frac{z_0}{z_1}
\label{RS}
\end{equation}
is the reduced source radius
which is close to the projected source radius onto the
screen plane when $z_1\gg z_0$, expressed in units of $\sqrt{\pi}R_F$.
It appears that
the diffraction patterns from the different parts of an extended source
will be washed out if $R_S\gg 1$ or, equivalently, if the angular radius
of the source $\theta_S$ is much larger than the angular Fresnel radius
$\theta_F$.
This is why there is a possibility to observe
a reasonably contrasted pattern only for extra-galactic sources, as
shown in table \ref{sources}. In this table, we have focused our
attention on the hottest
stellar classes that corresponds to the smallest stellar radii.

\begin{figure}
\begin{center}
\begin{turn}{0}
\mbox{\epsfig{file=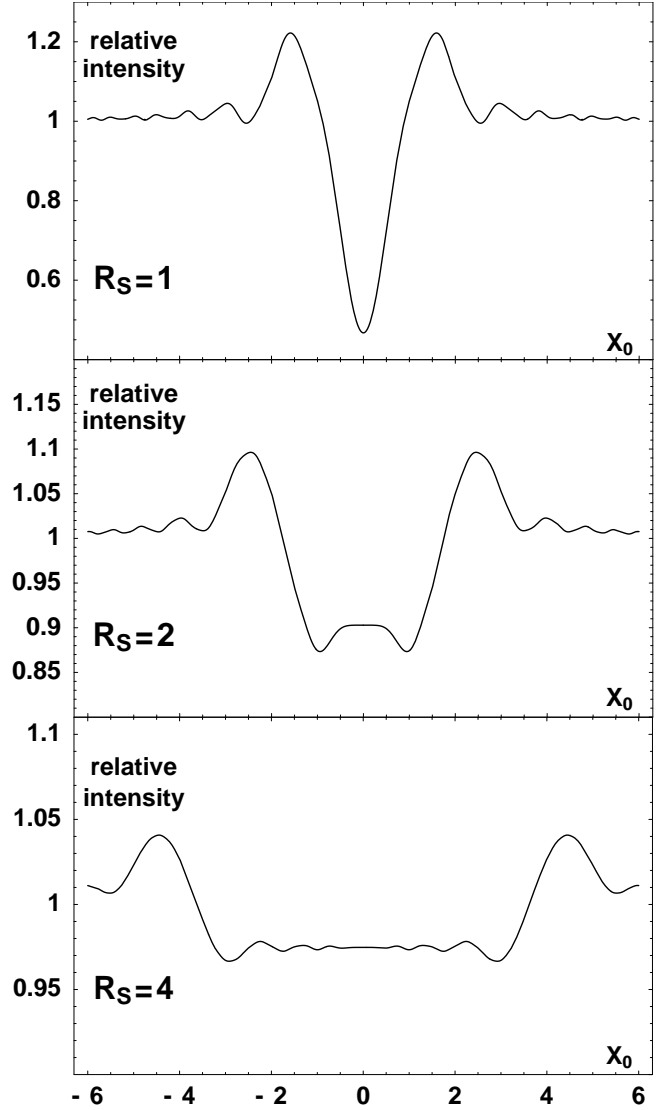,width=8.7cm}}
\end{turn}
\caption[]{Diffraction pattern
produced perpendicularly to a step of optical path
of $\delta=\lambda/2$, for extended disk-sources
of reduced radius $R_S=1$ (top), 2 (middle) and 4 (bottom).
The size of the central
depression corresponds to the reduced source size.
Notice that the vertical scales are very different because
the contrast is much
lower for $R_S=4$.
\label{diffstar}}
\end{center}
\end{figure}

\begin{figure}
\begin{center}
\begin{turn}{0}
\mbox{\epsfig{file=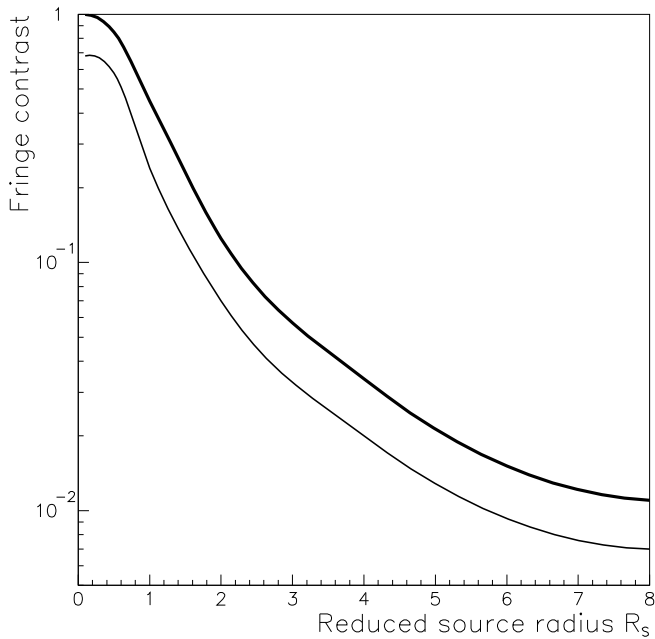,width=9.5cm}}
\end{turn}
\caption[]{Fringe contrast as a function of
the reduced star radius $R_S$ ($\sim \theta_S/\sqrt{\pi}\theta_F$
when the source is much farther from the observer than the screen),
for 2 values of the
step size: $\delta=\lambda/4$ (thin line) and
$\delta=\lambda/2$ (thick line). For a large source radius,
the contrast varies like $1/R_S$.
\label{hcontrast}}
\end{center}
\end{figure}

\begin{table*}
\begin{center}
\caption[]{Examples of sources at different distances.\\
- For LMC/SMC and M31 we have chosen
two series of sources with comparable apparent
luminosities ($M_V\sim 18.5$ and $\sim 20.5$).\\
- The size of a SNIa envelope is estimated assuming $10\,000\,\mathrm{km.s^{-1}}$ expansion rate during 20 days (\cite{Filippenko}).\\
- The angular and luminosity distances of the Einstein cross
multi-image quasar are estimated with the cosmological
parameters of \cite{WMAP}.
The upper size of the continuum source
is constrained by microlensing studies.
The absolute magnitude given here is only indicative, as it does not take into
account the magnification due to gravitational lensing.
}
\label{sources}
\begin{tabular}{|c|c|c|c|c|c|c||c|c|}
\hline
		&		& Distance	& Source or	& Absolute	& Apparent	&	& \multicolumn{2}{c|}{{\bf Angular size} $\theta_S$} \\
Source		& Distance	& modulus	& stellar type		& mag. V	& mag. V	& Size	& rad	& $\arcsec$ \\
\hline
Nearby star	& $10\,\mathrm{pc}$	& 0.		& M5V		& 12.3		& 12.3		& $0.27 r_{\odot}$	& $6.1.10^{-10}$	& $1.3.10^{-4}$ \\
Galactic star	& $8\,\mathrm{kpc}$	& 14.5		& K0V		& 5.9		& 20.4		& $0.85 r_{\odot}$	& $2.4.10^{-12}$	& $5.0.10^{-7}$ \\
\hline
LMC star	& $55\,\mathrm{kpc}$	& 18.7	& B8V	& -0.25	& 18.45	& $3.0 r_{\odot}$	& $1.2.10^{-12}$	& $2.5.10^{-7}$ \\
		&		&	& A5V	& 1.95	& 20.65	& $1.7 r_{\odot}$	& $7.10^{-13}$		& $1.4.10^{-7}$ \\
\hline
SMC star	& $58\,\mathrm{kpc}$	& 18.8	& B8V	& -0.25	& 18.55	& $3.0 r_{\odot}$	& $1.1.10^{-12}$	& $2.3.10^{-7}$ \\
		&		&	& A5V	& 1.95	& 20.75	& $1.7 r_{\odot}$	& $7.10^{-13}$		& $1.4.10^{-7}$ \\
\hline
M31 star	& $725\,\mathrm{kpc}$	& 24.3	& O5V	& -5.7	& 18.6	& $12.r_{\odot}$	& $3.7.10^{-13}$	& $7.6.10^{-8}$ \\
		&		&	& B0V	& -4.0	& 20.3	& $7.4r_{\odot}$	& $2.3.10^{-13}$	& $4.7.10^{-8}$ \\
\hline
\hline
SNIa@max & $0.9\,\mathrm{Gpc}$	& 39.8	& SNIa	&-19.2 & 20.6 & $\sim 2.10^{10} \,\mathrm{km}$ & $7.2.10^{-13}$ & $1.5.10^{-7}$ \\
at z=0.2 &			&	&	&      &      &				&		&	\\
\hline
Einstein cross A	& $1.75\,\mathrm{Gpc}$	& $45.5$	& quasar	& $\sim -28.$	& $\sim 17.$ & $<6.10^{10}\,\mathrm{km}^a$ & $<1.1.10^{-12}$ & $<2.3.10^{-7}$\\
z=1.695	& 		& 		& 		& 		& 	     & $<3.10^{11}\,\mathrm{km}^b$ & $<5.6.10^{-12}$ & $<1.1.10^{-6}$\\
\hline
\multicolumn{9}{l}{$^a$ \cite{Wyithe}.}\\
\multicolumn{9}{l}{$^b$ \cite{Atsunori}.}
\end{tabular}
\end{center}
\end{table*}
\begin{table*}
\begin{center}
\caption[]{Examples of screen positions, corresponding Fresnel sizes, and
fringing time scales at $\lambda=500\,\mathrm{nm}$.
The typical relative velocity of the nearby stars is taken from
the dispersions published by \cite{allen} (2000)}
\label{screens}
\begin{tabular}{|l|c|c|c||c|c|c||}
\hline
		&		& solar	& solar 	& Galactic & Galactic & Galactic \\
Screen type	& atmosphere	& system& neighbourhood  & thin disk& thick disk& halo \\
\hline
Distance	& $10\,\mathrm{km}$	& $1\,\mathrm{AU}$ & $10\,\mathrm{pc}$	& $300\,\mathrm{pc}$ & $1\,\mathrm{kpc}$ & $10\,\mathrm{kpc}$ \\
Fresnel size	& $2.8\,\mathrm{cm}$	& $109\,\mathrm{m}$& $157\,\mathrm{km}$	& $860\,\mathrm{km}$& $1570\,\mathrm{km}$ & $5000\,\mathrm{km}$ \\
\hline
{\bf Angular size} $\theta_F$ (rad.)	& $2.8.10^{-6}$ & $7.3.10^{-10}$ & $5.1.10^{-13}$ & $9.3.10^{-14}$& $5.1.10^{-14}$ & $1.6.10^{-14}$ \\
$\theta_F$ in $\arcsec$	& $0.6$ & $1.5.10^{-4}$ & $1.0.10^{-7}$ & $1.9.10^{-8}$& $1.0.10^{-8}$ & $3.3.10^{-9}$ \\
\hline
relative speed	& $1\,\mathrm{m/s}$	& $10\,\mathrm{km/s}$	& $20\,\mathrm{km/s}$ & $30\,\mathrm{km/s}$ & $40\,\mathrm{km/s}$ & $200\,\mathrm{km/s}$ \\
time scale	& $0.03\,\mathrm{s}$	& $0.01\,\mathrm{s}$	& $8\,\mathrm{s}$ & $29\,\mathrm{s}$ & $40\,\mathrm{s}$ & $25\,\mathrm{s}$ \\
\hline
\end{tabular}
\end{center}
\end{table*}
As an illustration, Fig. \ref{diffstar} shows the diffraction
patterns for a source of reduced radius equal to 1, 2 and 4.
Fig. \ref{hcontrast} gives the maximum contrast of the pattern
as a function of $R_S$ for two values of the step size $\delta$.
This shows
the very strict limitations on the angular size of the target object
to get a measurable contrast:
$\theta_S$ should not be too large with respect to $\theta_F$.

Examination of tables \ref{sources} and \ref{screens} shows that every star
gives a contrasted diffraction pattern through the atmosphere
if the optical path
changes by the order of $\lambda$ at the cm scale.
This happens for a star close to horizon, and explains the stronger
twinkling contrast observed at this moment\footnote{
When the stars are high in the sky, their twinkling originates in
the weak scintillation regime described below.}.
More interestingly, we note that only remote
stars have a chance to give a reasonably contrasted fringe system on
Earth through a Galactic screen (at $>300\,\mathrm{pc}$).
We will focus on two configurations:
A small source located in a Magellanic Cloud (LMC or SMC)
or a bigger source located in M31, with a screen
located at a typical halo object distance ($\sim 10\,\mathrm{kpc}$).

Numerical calculations show that the maximum contrast region
has the reduced size of the source (see Fig. \ref{diffstar}).
The maximum intensity occurs when
the source periphery is almost tangent to the step
when seen from the observer's plane.
When the step is out of the projected source disk,
the inter-fringe is $\simeq 1.5\times R_F$,
independently of the source radius.
\subsection{Chromatic source. Temporal coherence}
Up to now we have only considered monochromatic sources.
The standard UBVRI filter system has passbands that all satisfy
$\Delta \lambda/\lambda <0.1$.
With these filters, temporal coherence is sufficiently
high to enable the formation of contrasted interferences between
different parts of the wave-front,
as long as the optical path difference is less than
$\lambda^2/\Delta \lambda\sim 10\times\lambda$. As this coherence
length is large compared to the typical optical path differences
we consider here, we can ignore the temporal coherence
aspects in the following.

The inter-fringe scales with $R_F$, i.e. with $\sqrt{\lambda}$,
then $\Delta \lambda/\lambda <0.1 \Rightarrow \Delta R_F/R_F <0.05$.
The contribution of the wavelength dispersion
to the fringe jamming is then usually much smaller than the
contribution due to the source extension in the configurations
studied here.
\subsection{The observable phenomenon, summary}
Our study shows that an interference pattern with inter-fringe
of $\sim R_F$ ($100-10\,000\,\mathrm{km}$) is expected on Earth when
the line of sight of a sufficiently small astrophysical source
(such as a remote star) crosses the edge of a
structure that changes the optical path by a significant
fraction of $\lambda$.
Such structures move with respect to the line of sight with
typical velocities
given in table \ref{screens}.
The observer's plane is then illuminated by an interference
pattern moving with the same velocity (remember that the source is
much farther than the screen).
The shape of the interference pattern can also evolve, due to
random turbulence in the scattering medium. We will base the
present study on the assumption that the scintillation is
mainly due to pattern motion rather than the pattern instability,
as it is usually the case in radioastronomy observations (\cite{lyne} 1998).
The contrast of the pattern is critically limited by the
angular size of the source; in the configurations proposed
here, we expect a typical contrast ranging between 1\% and 10\%.
The time scale $t_{scint}$ of the intensity
fluctuations is $t_{scint} = R_F/V_T$, where $V_T$ is the transverse
velocity of the structure. It is of order of $10-40\,\mathrm{s}$.
As discussed just before, the inter-fringe
scales with $\sqrt{\lambda}$;
therefore, one expects a significant difference
in the inter-fringe and in the time scale $t_{scint}$
between the red side of the optical
spectrum and the blue side. This property
might be used to sign the diffraction
phenomenon at the $R_F$ natural scale. We will see below
that the $\lambda$ dependence is different if
the screen produces optical path stochastic fluctuations
at a length scale much smaller than $R_F$.
\section{General discussion of the scintillation regimes}
Up to now, we have considered variations of the screen optical
thickness at a scale close to $R_F$.
In the screen plane, $R_F$ can be considered as a coherence
domain, in the sense that -- as  mentioned earlier --
integral (\ref{amplit}) is dominated by the contribution of a region
characterized by the Fresnel radius. This means that only
the patch of the wavefront within a few Fresnel radii contributes
coherently to the integral.
The details of the optical path variations inside this domain
then drive the diffraction pattern, leading to the (small scale)
diffractive scintillation.
If there is a large scale structure ($\gg R_F$) in the optical path
variations, a succession of focusing and defocusing
configurations occurs, where the first phase term $k\delta(x_1,y_1)$
in integral (\ref{amplit}) can partially compensate or enhance the
second phase term; the
coherence domain that contributes to the integral can then
become larger or smaller than $R_F$.
In the case of large optical path variations,
several coherence domains can also converge
and the average luminosity can be estimated through
the geometrical optics approximation.
The intensity then depends on the local
focal length produced by the optical path variations on the screen.
Intensity variations -- called refractive scintillation -- arise at a
length scale larger than $R_F$.

Intensity scintillation is
well known in observations of compact radio sources.
Indeed scintillation of radio-pulsars is used to study the
nearby interstellar and solar system media.
The large wavelength and the
very small size of the sources allow one a good control of
this technique.
In this section, we will
summarize the conditions and characteristics for the
different scintillation regimes that occur in
pulsar radio-observations, and adapt them to star optical observations.
Extensive literature exists on pulsar physics.
Reviews can be found in \cite{lyne} (1998) and \cite{narayan92};
the scaling laws described below are taken and adapted for the latest
reference.

\begin{figure}
\begin{center}
\begin{turn}{0}
\mbox{\epsfig{file=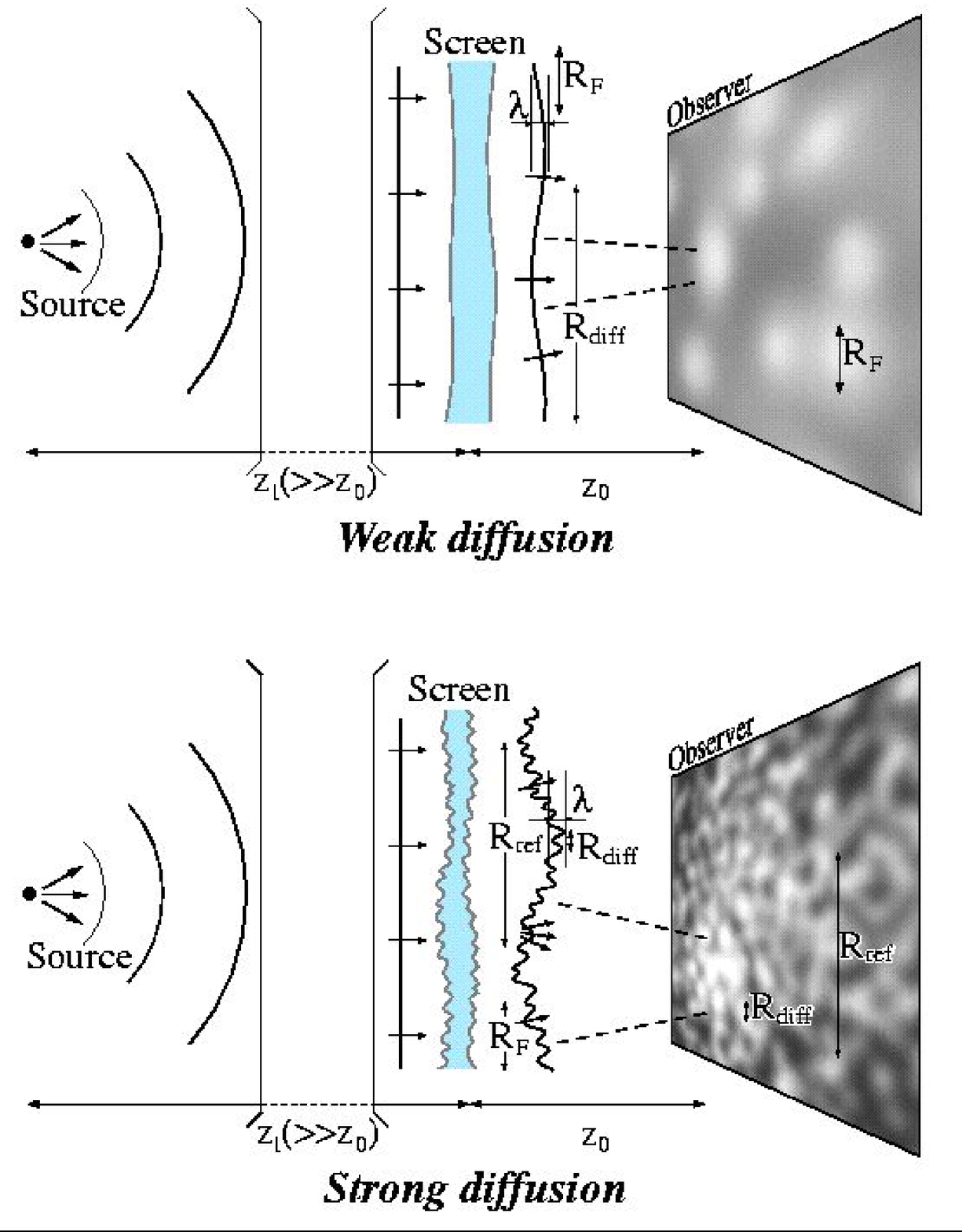,width=8.7cm}}
\end{turn}
\caption[]{The two scintillation regimes:
sections of the wavefront after screen crossing.\\
Top panel: $R_{diff}\gg R_F$. The weakly distorted
wavefront produces a weak scintillation at scale $R_F$
in the observer's plane.\\
Bottom panel: $R_{diff}\ll R_F$. The strongly distorted wavefront
produces strong scintillation at scales $R_{diff}$ (diffractive mode)
and $R_{ref}$ (refractive mode) in the observer's plane.
\label{front}}
\end{center}
\end{figure}

The distinction between the regimes is based on the relative values of the
length scale of the optical path fluctuations and of the Fresnel scale $R_F$.
The screen is characterized by its diffractive length scale
$R_{diff}$, defined as the separation in the screen plane
for which the root mean square of the optical path difference
$\delta(x'_1,y'_1)-\delta(x_1,y_1)$ is $\lambda/2\pi$ (see Fig. \ref{front}).
As the patch of the wavefront contributing to integral (\ref{amplit})
has a size of order of $R_F$ in the screen plane, we can
consider two very different situations since
integral (\ref{amplit}) has a completely different behaviour
whether $R_{diff}\gg R_F$ or $R_{diff}\ll R_F$.
\subsection{Weak scintillation regime}
If $R_{diff}\gg R_F$, then the random fluctuations of the optical path
are small compared to the wavelength within the Fresnel zone.
The screen is then qualified as weakly diffusive.
Only weak perturbations of the
wavefront are expected, essentially from the focusing/defocusing
effects of the optical path fluctuations {\it within} this Fresnel zone.
In this regime, only the wavefront patch enclosed in the Fresnel
radius plays a significant role.
Indeed as variations of $\delta(x_1,y_1)$
are small over the Fresnel domain, $\delta(x_1,y_1)$ cannot
significantly change the total phase in integral (\ref{amplit})
when $(x_1-x_0,y_1-y_0)$ is close to the Fresnel circle.

An alternative point of view is to consider that the
emerging wave is a combination of the initial plane wave plus weak
amplitude perturbing waves with a random distribution of wave normals.
For the observer, a point source located behind the screen would
appear like a main spot, surrounded by low luminosity spots.
The resulting intensity variations when the screen moves
from one Fresnel-size zone to the next
are weak, and their typical time scale is $R_F/V_T$, where $V_T$ is the
transverse velocity of the screen with respect to the observer-source
line of sight.
Table \ref{regimes} gives the conditions to observe
this scintillation regime and its characteristics.
\subsection{Strong scintillation regimes}
If $R_{diff}\ll R_F$, then the random fluctuations of the optical path
produce strong random phase changes within the Fresnel zone.
The screen is considered as a strongly diffusive filter because
the wavefront is strongly perturbed.
From our alternative point of view, one can consider that
the contribution of the initial plane
is almost completely redistributed after crossing the screen.
The point source seen by the observer through the screen
would now appear like scattered spots with random
intensities.
In this situation, the intensity
is expected to strongly vary
at the sub-Fresnel scale.
Moreover, large length-scale structures in the optical path variations
produce average focusing/defocusing effects over domains much larger than
the Fresnel size.
Two scintillation modes are then to be distinguished:
\begin{itemize}
\item
{\bf Diffractive scintillation} \\
\label{strongdiff}
As the optical path variations are larger than $\lambda$ within the
Fresnel radius, one expects a strongly contrasted fringe system
(for a point-like source).
By contrast with the simple case studied in Section \ref{simplecase},
where no explicit length scale other than $R_F$ was introduced,
there is now a different scale of optical path variations
$\delta(x_1,y_1)$ characterized by $R_{diff}$.
Then, when the screen is shifted from a position to
another located at distance $R_{diff}$,
the values of integral (\ref{amplit}) are not correlated between
the two positions.
The resulting intensity variations when the screen moves are
large, and their time scale is $R_{diff}/V_T$.

As discussed in Section \ref{spatialcoh},
diffractive scintillation is critically dependent
on the source size. In contrast,
the time scale does not depend on the wavelength $\lambda$,
but the {\it positions} of the maxima/minima do,
thus limiting the temporal coherence of the fringes.
When $\lambda$ changes by $\Delta \lambda$, the fringes
produced by $R_{diff}$-size structures are displaced by
$\sim z_0 \Delta \lambda/R_{diff}$. The fringe systems are decorrelated
when this displacement is $\sim R_{diff}$, i.e. when
$\Delta \lambda/\lambda\sim (R_{diff}/R_F)^2$ (see Table \ref{regimes}).
\label{largeurbande}
\item
{\bf Refractive scintillation} \\
This effect
results from the strong stochastic variations of the optical path
produced by large scale structures of the screen.
Let $R_{ref}$ be this additional length scale.
In contrast to the case of weak scintillation, the
large variations of the optical path can considerably change the
values of the integrand phase of integral (\ref{amplit}) at the limits
of the Fresnel domain. Focusing/defocusing effects are not
restricted to the Fresnel zone any more, and light from ``other''
Fresnel domains
can be focused if the optical path variations are large enough.
Such effects now involve regions of typical size defined by $R_{ref}$,
and their time scale is $R_{ref}/V_T$.

Source size is much less critical for the visibility of this
mode. The time scale also does not depend on the wavelength
and the illumination patterns are coherent within a wide
passband (see Table \ref{regimes}).
\end{itemize}
\begin{table*}
\begin{center}
\caption[]{Conditions and characteristics of the different
scintillation regimes for point-like and extended sources of
apparent angular size $\theta_S$.
$R_{scint}$ is the characteristic length scale
of the illumination pattern, and $t_{scint}$ is the characteristic time scale
of the intensity variations for a screen moving at transverse speed
$V_T$ with respect to the line of sight.
The modulation index $m_{scint}$ is the root mean square amplitude of the
intensity scintillation.
$m_{scint}$ depends only on the $\theta_{diff}$ to $\theta_S$ ratio
for the diffractive mode.
For the other scintillation modes, $m_{scint}$
is given for stochastic fluctuations of the
optical path length due to Kolmogorov turbulence in the scattering medium.}
\label{regimes}
\begin{tabular}{l|c|c|c|}
	& $R_{diff}\gg R_F$	& \multicolumn{2}{c|}{$R_{diff}\ll R_F$} \\
	& Weak scintillation	& \multicolumn{2}{c|}{Strong scintillation} \\
	&			& diffractive mode & refractive mode \\
\hline
characteristic length	& $R_F$	& $R_{diff}$	& $R_{ref}\gg R_{diff}$ \\
characteristic angle	& $\theta_F$	& $\theta_{diff}$ & $\theta_{ref}$ \\
coherence passband $\Delta \lambda/\lambda$ & $\sim 1$	& $(R_{diff}/R_F)^2$ & $\sim 1$ \\
\hline
Point-source condition	& $\theta_S<\theta_F$ & $\theta_S<\theta_{diff}$ & $\theta_S<\theta_{ref}$ \\
\hline
$R_{scint}$ & $R_F$ & $R_{diff}$ & $R_{ref}$ \\
$t_{scint}$ & $R_F/V_T$ & $R_{diff}/V_T$ & $R_{ref}/V_T$ \\
$m_{scint}$ & $(R_F/R_{diff})^{\frac{5}{6}}\ll 1$ & $\sim 1$ & $(R_{diff}/R_F)^{\frac{1}{3}}\ll 1$ \\
\hline
Extended source	& $\theta_S>\theta_F$ & $\theta_S>\theta_{diff}$ & $\theta_S>\theta_{ref}$ \\
\hline
$R_{scint}$ & $R_F(\theta_S/\theta_F)$ & $R_{diff}(\theta_S/\theta_{diff})$ & $R_{ref}(\theta_S/\theta_{ref})$ \\
$t_{scint}$ 	& $(R_F/V_T)(\theta_S/\theta_F)$ &
		$(R_{diff}/V_T)(\theta_S/\theta_{diff})$ &
		$(R_{ref}/V_T)(\theta_S/\theta_{ref})$ \\
$m_{scint}$ & 	$(R_F/R_{diff})^{\frac{5}{6}}(\theta_F/\theta_S)^{\frac{7}{6}}$ &
		$\theta_{diff}/\theta_S < 1$ &
		$(R_{diff}/R_F)^{\frac{1}{3}}(\theta_{ref}/\theta_S)^{\frac{7}{6}}$ \\
\hline
\end{tabular}
\end{center}
\end{table*}

\begin{figure}
\begin{center}
\begin{turn}{0}
\mbox{\epsfig{file=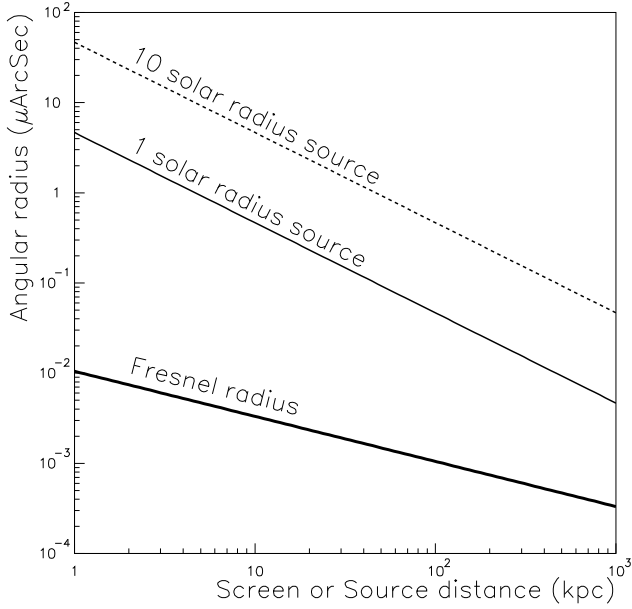,width=9.cm}}
\end{turn}
\caption[]{Apparent stellar angular radius $\theta_S$ for a solar-like
star ($r_s=r_{\odot}$) and for a star with $r_s=10r_{\odot}$
as a function of the distance.
The lower line gives the Fresnel radius $R_F$ as a function
of the screen's distance, illustrating the fact that the
source has to be much farther than the screen to give contrasted
diffractive scintillation.
\label{angles}}
\end{center}
\end{figure}

\begin{figure}
\begin{center}
\begin{turn}{0}
\mbox{\epsfig{file=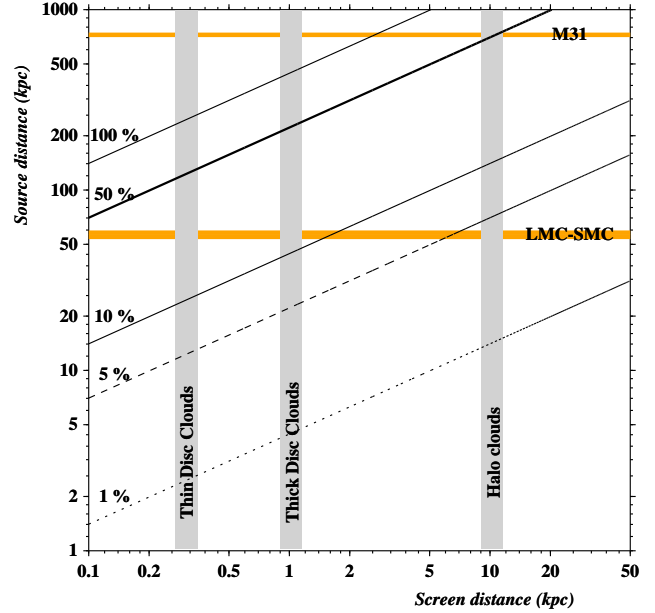,width=9.cm}
}
\end{turn}
\caption[]{Lines of equal diffractive scintillation modulation index
at $\lambda = 500\,\mathrm{nm}$ in the source/screen distances plane.\\
The modulation index is 100\% (thick line), 50\%, 10\%, 5\% and 1\% (from top
to bottom lines), assuming that the product
$\left[\frac{R_{diff}}{R_F}\right]\left[\frac{r_s}{r_{\odot}}\right]^{-1}$ is unity.
This happens only for very small sources, and when
$R_{diff}$ is not too small compared to $R_F$. Otherwise,
the modulation index has to be multiplied by this product.
Typical distances of disks and halo clouds are indicated,
as well as the distances of the two most promising targets.
\label{isocontraste}}
\end{center}
\end{figure}

The strong scattering diffraction mode is the most promising
in terms of contrast, as demonstrated in Section \ref{simplecase}.
Fig. \ref{angles} allows one to compare the apparent stellar angular
size $\theta_S$ with the natural angular Fresnel scale $\theta_F$ at
$\lambda=500\,\mathrm{nm}$.
Fig. \ref{isocontraste} shows the iso-contrast lines in the
fringing system expected in the case of strong diffractive scintillation.
Expressed in practical units, the modulation index as defined
in table \ref{regimes} is in this case
\begin{eqnarray}
\lefteqn{m_{scint}=} \\
& 0.071 \left[\frac{\lambda}{500\,\mathrm{nm}}\right]^{\frac{1}{2}}
\left[\frac{R_{diff}}{R_F}\right]\left[\frac{r_S}{r_{\odot}}\right]^{-1}
\left[\frac{z_0+z_1}{1\,\mathrm{kpc}}\right]\left[\frac{z_0}{1\,\mathrm{pc}}\right]^{-\frac{1}{2}}. \nonumber
\end{eqnarray}

The example discussed in Section \ref{simplecase} corresponds to a
diffractive scintillation regime that represents a
transition between the weak and the strong regimes. The natural
scale $R_F$ is here the only relevant parameter
depending only on the screen position and wavelength.
For this transition regime, the modulation index estimated
from the above
formula is stronger than calculated in Section \ref{simplecase},
because the approximations of Table \ref{regimes}
are not valid.
For a B0V star in M31 ($\theta_S=4.7.10^{-8}\ \arcsec$) and a screen at $1\,\mathrm{kpc}$
($\theta_F=1.0.10^{-8}\ \arcsec$), the expected contrast
for diffractive scintillation -- setting $R_{diff}=R_F$ in the
expression of Table \ref{regimes} -- is
$m_{scint}\sim (\theta_{diff}/\theta_S)\sim (\theta_F/\theta_S)=0.21$,
to be compared with the contrast of $\sim 0.1$ expected from a
step of $\lambda/2$ in the optical path (Fig. \ref{hcontrast}).
\subsection{Extreme scintillation}
For completeness, we should also mention an extreme scintillation
regime, due to caustic effects, occurring 
when the observer approaches the focus of a refractive structure.
Structures like the ones predicted by \cite{perspectives}, with
concentrations of diameter $D=10\,\mathrm{AU}$ that produce a total
variation in the optical path (from edge to center)
$\Delta=50\,000\times \lambda$ at $\lambda=500\,\mathrm{nm}$,
should act as optical lenses.
A naive focal length estimate gives $f\sim D^2/8\Delta=360\,\mathrm{Mpc}$.
The concentration power of such a lens is clearly much too weak to
give a measurable signal on Earth.
Moreover, the time scale of the intensity variations should be
$t_{scint}=D/V_T=(10\,\mathrm{AU})/(200\,\mathrm{km.s^{-1}})=87\,\mathrm{days}$.
Other authors have discussed in detail the possible caustic
effects with polytropic models of self-gravitating gas
clouds, and the consequences
on background star light-curves (\cite{draine},
\cite{rafikov}).
We will not consider this regime in the following.
\subsection{Regime discrimination}
The appropriate tool to connect the screen structure with the
scintillation features will be the temporal
power spectrum of the intensity fluctuations.
If strong diffractive and refractive regimes
take place together, two peaks are expected in the power
spectra. Their relative power
will also be connected with the source size.
The so-called inverse problem, i.e.
the determination of a density fluctuation model of the
interstellar medium from scintillation observations has been
extensively discussed by radio-astronomers (\cite{narayan88}),
but is beyond the scope of this exploratory paper.

We will mainly discuss here the perspectives of
the diffractive scintillation regime.
Nevertheless, we will also sometimes mention the refractive regime
because important density fluctuations may occur at a much
smaller scale than the global structure,
due to gaseous turbulence phenomena.
\section{Possible configurations for detectable diffractive scintillation}
\label{possconf}
One of the main outcomes of the previous discussions is that only
small angular sources may give a reasonably contrasted
diffractive scintillation.
We now focus on this diffractive process, because
its modulation index can be large and is easy to
predict with basically no hypothesis
on the detailed screen structure. We will only assume that a regime
characterized by $R_{diff}\le R_F$
can be established, or that at least a transitory regime as described
in Section \ref{simplecase} -- characterized by $R_F$ -- can occur,
for example if
an inhomogeneity due to a turbulent mechanism crosses the line of sight.
Table \ref{configurations} that  combines data from
Tables \ref{sources} and \ref{screens} lists some configurations
that should produce diffractive scintillation and gives their
characteristics.
Depending on the diffusion strength, other regimes may take place;
the reader may adapt the forthcoming discussion to these
regimes,
after rescaling the characteristic length, time and contrast according
to Table \ref{regimes}. Optical depth may also be considerably larger
in those other regimes because they are strongly dependent on the
screen structure. We will essentially consider minimum optical
depths in what follows.

To end the discussion about the screen configurations,
we have to mention here that a diffusive screen located near
the source, or in the galaxy of the source, do not produce
scintillation, due to the size of the source.
In such a configuration ($z_1\ll z_0$),
the approximation of an incident plane wave
on the screen is not valid any more, and calculation should be
redone starting with eq. (\ref{general}). The Fresnel zone is then defined
by $R_F=\sqrt{\lambda/2\pi (1/z_0+1/z_1)^{-1}} \sim \sqrt{\lambda z_1/2\pi}$
instead of $\sqrt{\lambda z_0/2\pi}$ (see e.g. \cite{born} or \cite{sommerfeld}).
The $z_0/z_1$ multiplier that enters
the reduced radius expression (\ref{RS}) is very large;
consequently, spatial coherence is completely lost
in this situation.
Moreover, a diffusive screen {\it close} to the source cannot increase
significantly its apparent size for a simple geometrical reason:
the image of a source is not significantly distorted by diffusion occurring
near the emission point. We will ignore such effects
in the following.
\begin{table*}
\begin{center}
\caption[]{Configurations leading to strong diffractive scintillation.
Numbers are given for $\lambda=500\,\mathrm{nm}$. Interstellar absorption is
not taken into account for the magnitude estimates.
}
\label{configurations}
\begin{tabular}{|c|c|c|c||c|c|c||c|c|c|}
\cline{5-10}
\multicolumn{4}{c||}{} & \multicolumn{6}{c|}{\bf SCREEN} \\
\cline{5-10}
\multicolumn{4}{c||}{} & atmos- & solar	& solar neigh-	& Galactic & Galactic & Galactic \\
\multicolumn{4}{c||}{} & phere & system& bourhood  & thin disk& thick disk& halo \\
\cline{3-10}
\multicolumn{4}{r||}{Distance}	& $10\,\mathrm{km}$	& $1\,\mathrm{AU}$ & $10\,\mathrm{pc}$	& $300\,\mathrm{pc}$ & $1\,\mathrm{kpc}$ & $10\,\mathrm{kpc}$ \\
\cline{2-10}
\multicolumn{4}{r||}{$R_F$ to multiply by $\left[\frac{\lambda}{500\,\mathrm{nm}}\right]^{\frac{1}{2}}$} & $2.8\,\mathrm{cm}$	& $109\,\mathrm{m}$& $157\,\mathrm{km}$	& $860\,\mathrm{km}$& $1570\,\mathrm{km}$ & $5000\,\mathrm{km}$ \\
\cline{1-10}
\multicolumn{4}{r||}{$t_{scint}$
to multiply by $\left[\frac{\lambda}{500\,\mathrm{nm}}\right]^{\frac{1}{2}}
\left[\frac{R_{diff}}{R_F}\right]$}	& $0.03\,\mathrm{s}$ & $0.01\,\mathrm{s}$ & $8\,\mathrm{s}$ & $29\,\mathrm{s}$ & $40\,\mathrm{s}$ & $25\,\mathrm{s}$ \\
\cline{1-10}
\multicolumn{4}{r||}{Optical depth $\tau_{scint}$} & $1$ & & & \multicolumn{3}{c|}{total $>10^{-7}$} \\
\hline
\multicolumn{4}{r||}{$m_{scint}$ in \% to multiply by} & & & & & & \\
\multicolumn{4}{r||}{
$\left[\frac{\lambda}{500\,\mathrm{nm}}\right]^{\frac{1}{2}}
\left[\frac{R_{diff}}{R_F}\right]\left[\frac{r_S}{r_{\odot}}\right]^{-1}$
}	& $\sim 100\%$	& $32\left[\frac{d}{10\mathrm{pc}}\right]$ & $2.2\left[\frac{d}{1\mathrm{kpc}}\right]$	& $4.1\left[\frac{d}{10\mathrm{kpc}}\right]$	& $2.2\left[\frac{d}{10\mathrm{kpc}}\right]$ & $7.1\left[\frac{d}{100\mathrm{kpc}}\right]$ \\
\hline
\hline
\multicolumn{4}{|c||}{\bf SOURCE (\cite{allen} 2000)} & \multicolumn{6}{c|}{\bf DIFFRACTIVE MODULATION INDEX $m_{scint}$} \\
\cline{1-4}
Location & Type & $r_s$ & $M_V$ & \multicolumn{6}{c|}{\bf (to multiply by $\sqrt{\lambda/500\,\mathrm{nm}}\times R_{diff}/R_F$)} \\
\hline
\cline{5-10}
nearby	& B8V & $3.r_{\odot}$  & -.25 &  & 10\% & & & & \\
star	& A5V & $1.7r_{\odot}$  & 1.95 &  & 20\% & & & & \\
$d=10\,\mathrm{pc}$ & K0V & $0.85r_{\odot}$  & 5.9 &  & 40\% & & & & \\
	& M5V & $0.27r_{\odot}$  & 12.3 &   & $\sim 100\%$ & 1\% & & & \\
\cline{1-4}\cline{6-10}
Polaris	& F7I & $100r_{\odot}$  & 1.97 & $\ll 1\%$ & 4\% & & & & \\
$d=130\,\mathrm{pc}$& & & & & & & & & \\
\cline{1-4}\cline{6-10}
	& K0III & $15.r_{\odot}$ & 15.2 &  & 100\% & 1\% &  &  &  \\
Galactic & G5III & $10.r_{\odot}$ & 15.4 &  & 100\% & 2\% &  &  &  \\
star	& B5V & $3.9r_{\odot}$ & 13.3 & for & 100\% & 5\% & 1\% &  &  \\
	& A5V & $1.7r_{\odot}$ & 16.5 &  & 100\% & 10\% & 2\% &  &  \\
$d=8\,\mathrm{kpc}$ & K0V & $0.85r_{\odot}$ & 20.4 & every & 100\% & 20\% & 4\% &  &  \\
	& M5V & $0.27r_{\odot}$  & 26.8 &  & 100\% & 65\% & 12\% & & \\
\cline{1-4}\cline{6-10}
	& K5III & $25.r_{\odot}$  & 18.5 & source & 100\% & 5\% & 1\% & & \\
LMC	& O5V & $12.r_{\odot}$  & 13. &  & 100\% & 10\% & 2\% & 1\% & \\
	& B8V & $3.r_{\odot}$  & 18.5 & in a & 100\% & 40\% & 8\% & 4\% & 1\% \\
$d=55\,\mathrm{kpc}$ & A5V & $1.7r_{\odot}$  & 20.7 &  & 100\% & 70\% & 13\% & 7\% & 2\% \\
	& K0V & $0.85r_{\odot}$  & 24.6 & telescope & 100\% & 100\% & 27\% & 14\% & 5\% \\
\cline{1-4}\cline{6-10}
	& B5I & $50.r_{\odot}$  & 18.1 & & 100\% & 32\% & 6\% & 3\% & 1\% \\
M31	& O5V & $12.r_{\odot}$  & 18.6 & $>1\,\mathrm{m}$ & 100\% & 100\% & 25\% & 13\% & 4\% \\
$d=725\,\mathrm{kpc}$	& B0V & $7.4r_{\odot}$  & 20.3 &  & 100\% & 100\% & 40\% & 22\% & 7\% \\
	& B8V & $3.r_{\odot}$  & 24. & (see text) & 100\% & 100\% & $\sim 100\%$ & 53\% & 17\% \\
\cline{1-4}\cline{6-10}
\cline{1-4}\cline{6-10}
z=0.2	& SNIa & $2.10^{10}$ & 20.6 &  & 100\% & 70\% & 13\% & 7\% & 2\%\\
$d=0.9\,\mathrm{Gpc}$ & & $\mathrm{km}$ & & & & & & & \\
\cline{1-4}\cline{6-10}
Einstein & Quasar & $<6.10^{10}$ & $\sim 17$ &  & 100\% & $>45\%$ & $>8\%$ & $>4\%$ & $>1.4\%$\\
cross & Q2237 & $\mathrm{km}$ & & & & & & & \\
$d=1.75\,\mathrm{Gpc}$ & +0305 & $<3.10^{11}$ & $\sim 17$ &  & 100\% & $>9\%$ & $>2\%$ & $>1\%$ & $>0.3\%$\\
\hline\end{tabular}
\end{center}
\end{table*}
\section{Optical depths and event rates for diffractive and refractive scintillations}
In this section, we want to quantify the probability to
observe scintillation produced by disk or halo molecular
clouds.
The $1\%$ surface filling factor predicted in the model of \cite{fractal}
for gaseous structures 
is also the maximum optical depth for all the
possible refractive (weak or strong) and diffractive scintillation regimes.
Nevertheless, we want to consider here the pessimistic case where the
optical depth for the strong regimes is much smaller:
in the thick disk,
the typical transverse speed of a structure is $40\,\mathrm{km.s^{-1}}$.
Then the typical crossing time of such a structure should be
$\sim 400$ days.
Under the hypothesis that strong diffractive regime is expected
only when the structure enters or leaves the line of sight\footnote{The
validity of this hypothesis as well as the optical depth
estimate for each scintillation regime depend on the internal structure
model. Turbulence or any process creating filaments, cells, bubbles
or fluffy structures should be taken into account.},
the duration for this regime is of order of
$\sim 5$ minutes (time to cross a few fringes).
Then the optical depth $\tau_{scint}$ for such
regime is at least of order of $10^{-7}$ and the average exposure
needed to observe one event of $\sim 5$ minute duration
is $10^{6}\,\mathrm{star\times hr}$.

If the gaseous structures belong to the Galactic halo instead of the
thick disk, the same order of magnitude is also expected for the
optical depth.

We will neglect the multi-diffusion eventuality, given the small sky
fraction occupied by the structures. 
For our study, the case of a thick (longitudinally extended) screen
needs a formalism adaptation only if the extension along the line
of sight is such that $R_F$ significantly changes within the structure.
Several authors (\cite{lyne} 1998, \cite{tatarskii}) have
studied very thick screens and found a similar behaviour
to the one that prevails for thin screens.
\section{Feasibility of observations}
The diffractive regime with Galactic hidden $\mathrm{H_2}$ gas
(see the three last columns of Tables \ref{screens} and \ref{configurations})
can happen with contrasts better than 1\% only
if the source has an angular size smaller than a few $\sim 10^{-12}\,\mathrm{rad}$.
For a given surface temperature this constraint is
equivalent to a constraint put on the source
(star, SN, quasar...) magnitude. Choosing objects with the highest
surface luminosity in Table \ref{configurations} indicates that the
minimal magnitude of stars
whose light is likely to undergo a few percent modulation
index is about $M_V=20.5$.
Therefore, diffractive scintillation search needs the capability to sample
every $\sim 10\,\mathrm{s}$ (or faster) the luminosity of stars with $M_V>20.5$,
with a point-to-point precision better than a few percent.
This performance can be
achieved using a 2 meter telescope with a high quantum efficiency detector.

Indeed, the optimal relative photometric precision $\Delta$ is related
to the star magnitude $M$, the exposure time $T_{exp}$, the seeing,
the sky background magnitude $\mu$ per $\mathrm{arcsec^2}$ and the
telescope collecting surface $S_{tel}=\pi D_{tel}^2/4$ through the following formula,
taken from (\cite{SNmarly}):
\begin{eqnarray}
\label{tpose}
T_{exp} & \simeq & \frac{10^{M/2.5}}{S_{tel}\Phi(M=0)}
\frac{1}{\Delta^2} \\
 & \times & \left[0.27+1.5\sqrt{0.25+1.13\left[\frac{seeing}{1\ \arcsec}\right]^2
10^{\frac{M-\mu}{2.5}}}\right]^2, \nonumber
\end{eqnarray}
where $\Phi(M=0)$ is the photo-electron flux
collected at observatory level by the detector in the considered passband
per $\mathrm{m^2}$ collecting surface, for a $M=0$ star.
This relation is valid as long as the readout noise remains negligible.
For a telescope with two reflectors of 0.87 transmission each,
equipped with a high quantum efficiency detector ($\sim 90\%$),
$\Phi(M_V=0)\sim 4.2\times 10^9 \gamma \mathrm{e^-/s/m^2}$ in the V-band.
For good observation conditions ($seeing=1 \arcsec$, $\mu>21.7$, i.e. moon-less
nights),
the squared term of expression (\ref{tpose}) is $\sim 2$ for stars
with $M_V=20.5$, and negligibly smaller for somewhat brighter stars.
Then the relation simplifies into
\begin{equation}
\Delta \simeq 30\% \times 10^{\frac{M_V-20.5}{5}}\times \frac{1}{\sqrt{T_{exp}\,\mathrm{(s)}}}
\times \frac{1}{D_{tel}\,\mathrm{(m)}},
\end{equation}
where $D_{tel}$ is the diameter of the telescope.
The photometric precision $\Delta$ would be $5\%$ with $10\,\mathrm{s}$ exposures
on a $M_V=20.5$ star, with a 2 meter telescope.
Neglecting the interstellar absorption, this is
sufficient to detect the most contrasted (and the longest)
diffractive scintillation events,
as can be seen in Table \ref{configurations}.

We wish to stress that
the continuity of the monitoring is {\it not} a critical issue.
As events last only for a few
minutes in our scenario, the detection efficiency will not
be affected by data taking interruptions, contrary to the case of
microlensing searches.
The essential parameters for the sensitivity of a detection
setup will be the integrated exposure in $\mathrm{star\times hr}$\footnote{If
dead-times larger than a few minutes occur, they should
be subtracted from the exposure
because events can be contained and subsequently lost in such
time intervals. This is also a noticeable difference from
the microlensing searches.}, the
sampling rate and the photometric precision.

The variation of the diffractive scintillation pattern with
the wavelength $\lambda$ should also be used as a signature of the
process. 
As emphasised in Section \ref{strongdiff}, if $R_{diff}\ll R_F$
the time scale does not depend on $\lambda$, but the phase
(the timing of extrema) changes with $\lambda$.
In the particular case where $R_{diff}\sim R_F$,
the peak of the temporal power spectrum is expected
to scale with $\sqrt{\lambda}$.
For these reasons, multi-wavelength detection capability
is highly desirable. The most powerful approach would be to
get star spectra at a high time sampling rate.
The equivalent in radio-astronomy is called the dynamic spectrum,
showing the intensity fluctuations in a 2D (time versus wavelength)
diagram (see for example \cite{gupta}). 

According to Table \ref{configurations}, M31 hot main sequence
stars are the most promising targets, but LMC and SMC small stars
are probably easier to distinguish from their surroundings.
\section{Foreground effects, background to the signal}
One of the challenges to extract an {\it interstellar} intensity
scintillation signal is the disentangling of the scintillation
due to the
foreground media (atmosphere, solar system and nearby medium).
\subsection{Atmospheric intensity scintillation and absorption}
Atmospheric intensity scintillation should be
easier to handle than naively expected.
\cite{dravins} report extensive studies of this
phenomenon, that is connected with seeing studies.
When short length scale interference patterns (a few cm)
produced in the diffractive mode enter a
large aperture telescope, the collected light is averaged,
since the aperture acts like a low pass filter.
The modulation index is then much smaller than 1\% for
a telescope diameter larger than one meter.
Moreover, the time scale of this diffractive regime is two
orders of magnitude faster than the searched Galactic signal.
As far as the weak intensity scintillation is concerned, it also
induces very small modulation index $m_{scint}$
at $1-10\,\mathrm{s}$ time scale; the following formula
gives the low-frequency component of intensity scintillation at optical
wavelengths, at altitude site $h$ (\cite{dravins} part III):
\begin{equation}
m_{scint}(t_{exp}) \sim 0.09\ \frac{D_{tel}^{-\frac{2}{3}}}{\sqrt{2t_{exp}}}\ \cos(Z)^{-1.75}\ e^{\frac{-h}{8000\,\mathrm{m}}},
\end{equation}
where $\cos(Z)^{-1}$ is the airmass and $t_{exp}$ is the integration
time.
For a 2 meter telescope, an airmass of 2 and an observatory altitude
of $2400\,\mathrm{m}$, we find that $m_{scint}(1\,\mathrm{s})=3.10^{-3}$ and
$m_{scint}(10\,\mathrm{s})=10^{-3}$. Thus intensity variations
due to atmosphere will not seriously affect a diffractive interstellar
scintillation signal of $1\%$ amplitude or more.

Any other long time scale atmospheric effects such as
absorption changes at the sub-minute scale
(due to fast cirruses for example) should be easy to remove as long
as nearby stars are monitored together.
At this time scale, the light of stars within a small angular distance
undergo the same atmospheric effects at the same time, or with a
short delay. A careful subtraction of such collective effects
should in principle considerably reduce this source of background.
\subsection{Intensity scintillation due to the solar system medium,
and to solar neighbourhood structures}
\begin{itemize}
\item
{\bf Interplanetary medium: from $0$ to $5.10^{-6} \,\mathrm{pc}$}\\
As can be seen Table \ref{configurations}, diffractive scintillation
produced by interplanetary medium have a characteristic time
of $10^{-2}\,\mathrm{s}$, therefore it does not disturb an hypothetic
Galactic gas signal.

The eventuality of a slow refractive regime has
to be examined in more details:
Studies of the atmospheric properties by (\cite{dravins} part I)
were mainly done with $Polaris$ star. This
star is far enough ($130 \,\mathrm{pc}$) to scintillate through interplanetary
screens. Unfortunately, this star is big ($\sim 100 r_{\odot}$)
and its light cannot be subjected to modulations of
more than $4\%$ in the diffractive
regime\footnote{$Polaris$ is also a spectroscopic binary system that
may complicate the interference system.}.
Nevertheless, 
in the hypothesis of a refractive regime that we are discussing here,
we know that this not very critical, and
{\it Polaris} measurements should then really be a reliable indicator
of this latter scintillation type.
As the measured modulation index of this star
is much smaller than $1\%$ at any frequency, this gives us a good
preliminary indication that both diffractive and refractive
scintillations from interplanetary medium are probably negligible
toward {\it Polaris} direction.
\item
{\bf Solar neighbourhood: from $5.10^{-6}$ to $10 \,\mathrm{pc}$}\\
This conclusion also applies to solar neighbourhood screens
(up to $\sim 10 \,\mathrm{pc}$) for the {\it refractive} regime, but
the size of $Polaris$ prevents us from drawing any conclusion relative to
the {\it diffractive} regime.

The solar system is embedded in a local interstellar cloud extending
not farther than $10-20\,\mathrm{pc}$. The average column density of
atomic hydrogen through this structure is less than
$10^{19}\,\mathrm{cm^{-2}}$ towards the Galactic center
(\cite{LIC}, \cite{Ferlet}),
which is negligible compared with
the column density of the molecular clouds we are looking for.
If molecular or atomic overdensities occur in this local region
-- typically at $10\,\mathrm{pc}$ --, then according to
table \ref{configurations},
many types of stars located at $\sim 8\,\mathrm{kpc}$,
including red giants, should undergo
a contrasted diffractive scintillation;
as it can be deduced from this table \ref{configurations},
the distinctive feature of scintillation through more distant
screens ($>300\,\mathrm{pc}$) is that {\it only the smallest stars}
are expected to scintillate.
It follows that a strategy consisting in the simultaneous monitoring of
many different types of stars located at different distances
should allow one to discriminate effects due to solar neighbourhood
gas and due to more distant gaseous structures.
\item
{\bf The local interstellar medium: from $10$ to $200 \,\mathrm{pc}$}\\
The local interstellar cloud itself is located inside a bubble,
which is a $50-200\,\mathrm{pc}$ cavity mainly containing hot ionized gas.
Ionized hydrogen has a negligible effect on the visible light
propagation. The remaining atomic plus molecular hydrogen components
contribute for a total column density of less
than $~10^{20}\,\mathrm{cm^{-2}}$ up to $200\,\mathrm{pc}$ (\cite{Lehner}, \cite{Lallement}).
This is again much smaller than the
column density due to molecular clouds.
If dense atomic or molecular
structures are located at such distances or beyond,
they start to enter our domain of interest.
\end{itemize}
\subsection{Possible sources of fake signal?}
\label{bruit}
There is no known physical process that can produce $>1\%$ intrinsic
intensity variations of an ordinary star at the minute time scale.
Asterosismology involves
acoustic modes that produce a few tens of $ppm$ intensity variations
in the ``high'' frequency domain of a few minutes (\cite{Christensen}).
Planet transit could also give a fast luminosity change, but
are also expected to exhibit intensity variations smaller
than $0.1\%$.
Granularity of the star surface, spots or eruptions would induce
much lower frequency intensity variations than the diffractive
scintillation.
A few categories of recurrent variable stars exhibit important emission
variations at the minute time scale (\cite{sterken}). Among them are the
rare types UV Ceti and flaring Wolf-Rayet stars.
Both types are easy to identify from their spectral characteristics.
UV-Ceti are also very faint stars (absolute magnitude $>15$) and
only the closest ones could contaminate a monitoring sample.
\subsection{Technical limitations}
Blending of stars, which is common in the LMC/SMC/M31
crowded fields, may attenuate the measured contrast of a
fringed pattern.
Careful simulations are needed to measure the impact of these
limitations. Adaptative optics or space measurements are
possible ways to reduce this impact.\\
Another technical difficulty is the risk of getting
complicated (and fluctuating)
point spread functions as a result of the very short
exposures. Such circumstances could make necessary the use of
performant subtraction algorithm, given the very crowded
environment.
\section{Observation specifications and strategy}
A careful feasibility study should be made before starting an
ambitious observation program:
The stability of light curves of nearby stars should be checked
with a low cost telescope at the sub-minute time scale,
to test the control of collective intensity fluctuations
due to meteorological phenomena and to
explore possible limitations due to interplanetary gas.
After such preliminary studies, an observational program could
start with the specifications described below.
These specifications are optimized for the search for
diffractive scintillation in a large sample of stars.
We showed in table \ref{configurations}
that supernov{\ae} and ``small'' quasars\footnote{
The Einstein's cross Q2237+0305 considered in the tables
is the quasar that has the most precise size determination.
It is also located at an optimal redshift
z=1.695, corresponding to the maximum possible angular
distance according to the currently published cosmological
parameters, and then to a minimal angular size $\theta_S$.
}
may also be subjected to diffractive scintillation;
but the possible smallness of the optical depth ($\sim 10^{-7}$)
and the scarcity of such sources probably don't make them
suitable targets for the first searches of diffractive scintillation.
Moreover, one should not forget that diffractive scintillation
of such sources is not expected from clouds located far beyond the Galaxy
(see end of Sect. \ref{possconf}).  
Conversely, refractive regimes, which may have much larger optical depth,
could be searched on any target.
\subsection{From Earth, with a single telescope}
\begin{itemize}
\item
Optics

As already mentioned, at least a 2 meter class telescope is required
to search for scintillation in $<10\,\mathrm{s}$ exposures
on $20<M_V<21$ LMC/SMC or M31 stars.

First order adaptative optics will lead to
a better photometric precision,
but second-order adaptative optics (that corrects focusing/defocusing
effects of atmosphere) is probably not really necessary since
the atmospheric intensity scintillation should be negligible.
\item
Filters

Monitoring in infra-red wavelength is not clearly a decisive advantage;
the Fresnel radius is larger, but red stars are also usually
larger. In contrast, monitoring stars with at least
two different passbands should help to disentangle diffractive
scintillation from the refractive one. In the
case of diffractive scintillation, the phase differences
of the fringe systems between two different passbands will provide
constraints on the ratio of the screen distance to the diffusion
scale $R_{diff}$ (see Section \ref{largeurbande}).
\item
Detector

A fast, low-background readout detector is essential to make possible
the requested high sampling rate and to maintain a decent useful to
dead time ratio.
A possible option could to put an array of small (high quantum
efficiency) frame-transfer CCDs at the focus of the telescope, that
could be rapidly read (because they are small) in parallel during the
exposures.

An alternative could be to use a series of very narrow CCDs (short columns),
with continuous slow readout of the lines, to collect continuous
light-curves without any dead-time.
Such continuous readout option is not viable
with a normal (not narrow) CCD, because of the
light-curves mixing due to field crowding.
\item
Targets and fields

Imaging $1\,\mathrm{deg^2}$ field on isophot $M_V=23\,\mathrm{mag/arcsec^2}$
towards LMC/SMC/M31 allows one
to monitor of order of $10^5$ ($20<M_V<21$) stars\footnote{
In more luminous fields like in the LMC bar, these stars
would be blended by more luminous ones.} (\cite{Elson}; \cite{hardy}).
It is essential to choose fields
containing stars with radii spanning a wide interval,
and also containing a sample of nearby stars.
This is necessary
to extract information about the screens through the
relationship between the temporal power
spectrum and the source size and distance.
In particular, nearby stars should provide a control
sample of non-scintillating light-curves.
\item
Telescope time

Exposure of a few $10^7\,\mathrm{star\times hr}$ (on small stars
with $20<M_V<21$) during dark nights
would provide a significant sensitivity to
the existence of Galactic $\mathrm{H_2}$ structures.
This needs typically one season of
observations using a wide field telescope ($1\,\mathrm{deg^2}$)
or a few seasons with a standard field telescope.
Nevertheless, as
already mentioned, there is no need for consecutive
telescope time, making such program relatively flexible.
Every sequence of consecutive images
-- longer than a few minutes -- will contain its own
events and will then be autonomous.
Distinct but possibly simultaneous campaigns
with different air-masses could certainly improve
the knowledge of atmospheric effects.
\item
Complementary observations

In case of positive detection, complementary observations should be
planned for scintillating candidate objects.
The first requirement is the identification of the stellar type
through spectroscopy, in order to check that
the candidate do not belong to the very specific cataclysmic objects
mentioned in section \ref{bruit}, and to get an estimate of the
distance and radius of the candidate. Such estimates are necessary
to constrain the maximum distance of the screen.

In the case of a star appearing to scintillate for a long time,
time-resolved spectroscopy or multi-wavelength data taking as well as
multiple detection (see below) will allow one to check the hypothesis
of diffractive or refractive scintillation.
\end{itemize}
\subsection{From Earth: array of telescopes}
A 2D array of telescopes, a few hundred
and/or thousand kilometers apart, would
certainly be the most powerful system to
measure the characteristics of a diffusive screen:
One single telescope will only provide a degenerated information
(the time scale $t_{scint}=R_{diff}/V_T$). Sampling a diffraction pattern
with a 2D array would allow one to separately measure
the geometrical scale $R_{diff}$ and the speed of the pattern.
Atmospheric effects will be decorrelated between the telescopes,
as well as the interplanetary gas effects. Nearby gaseous structures
($\sim 10\,\mathrm{pc}$) are expected to produce $\sim 150\,\mathrm{km}$ fringes
on Earth and Galactic structures are expected to 
produce $>1000\,\mathrm{km}$ fringes, that should be easy to distinguish
with a handful of synchronized telescopes sampling
a few thousand kilometers wide pattern.
The speed and direction of the diffraction pattern
(drift due to the relative velocity of the screen with respect to the
line of sight), as well as the pattern's variations
(due to the dynamics of the scattering medium), should also be
measurable through the analysis of time delay between the telescopes.
Obviously, any intrinsic star variability will be unambiguously
identified, as it does not produce an intensity geometrical pattern.
\subsection{From space}
The main advantages of a space mission
would be the better photometric precision
and especially the much better spatial resolution, allowing to
seriously reduce blending problems.
As the intensity scintillation due to
atmosphere is not critical at the level of $1\%$, the
cancellation of this specific scintillation in
a spatial observatory will be interesting only for
very weak scintillation searches.
Among the operational spatial observatories and the planned missions,
the HST observatory and COROT mission have been examined:
\begin{itemize}
\item
COROT mission (\cite{COROT}) plans to monitor stars with a short time-scale
sampling, but $1\%$ precision photometry on $M_V=20$
stars in $10\,\mathrm{s}$ exposures is definitely unaccessible due to the modest size
of the telescope.
\item
The Space Telescope Imaging Spectrograph instrument (STIS)
of the Hubble Space Telescope (HST) is the only instrument that is
able to offer a cycle time of $\sim 20\,\mathrm{s}$ for time-resolved
spectroscopy or photometry; but this mode is only available on
a sub-array of the CCD (\cite{STIS}). Regrettably, only one
(very) wide band filter
($550-1100\,\mathrm{nm}$) can be used to get $\sim 5\%$ photometric precision
in $10\,\mathrm{s}$ exposure time on a $M_V=20.5$ B0V star. This very wide
band filter would already limit the temporal coherence, thus decreasing
the diffractive scintillation contrast.
The other filters have too narrow passbands to enable short exposures.
Nevertheless, the HST angular resolution would be extremely
valuable for monitoring M31 fields, and there could be an
interesting opportunity to further investigate.
The other HST instruments have prohibitive overhead times.
\end{itemize}
\section{Data flow and analysis}
Data flow and analysis problems can be considered as the result of
an hybrid of the EROS survey -- for the massive photometric reduction --
and the VIRGO gravitational wave experiment.
For example, robust filtering developed in VIRGO for burst searches
(\cite{VIRGO}) should be appropriate to
the search for an oscillating signal within a definite short period.
The time scales of the data flow and of the signal are just multiplied by five
orders of magnitude with respect to VIRGO.
\section{Studies to be done, developments and perspectives}
A simulation of the Galactic $\mathrm{H_2}$ gas distribution, and
-- more important -- of the intra-cloud turbulence is clearly needed
to produce more quantitative predictions on the optical depth, on the
signal shape and on the temporal power spectrum.

Examination of the inverse problem will lead to more precise ideas
on the information about the screen and the source
that could be extracted from a statistical
analysis based on the temporal power spectra.
For example, as there is a clear connection between the diffractive
pattern, the fringe contrast
and the source size, the scintillation process may
also be useful to improve the stellar radii knowledge, and to
constrain supernov{\ae} and quasar dimensions.

It has been amply demonstrated that the angular size of the source
is the critical parameter for the fringe contrast.
Observations with a very large telescope would make possible the monitoring
of stars much more distant than M31, thus providing more contrasted
diffractive intensity scintillation effects.

Existing data sets may
allow the extraction of interesting constraints
on the {\it refractive} scintillation mode. For instance
EROS1 experiment produced $20\,000$ photometric measurements
of $\sim 100\,000$ stars in the LMC bar (\cite{EROS1}).
12 minutes blue and 8 minutes red exposures were alternatively
taken.
As refractive scintillation mode
is much less sensitive to the source size
than diffractive scintillation mode, the sources monitored by EROS1
were likely to scintillate.
As this refractive mode could produce long time scale
variations (longer than a few minutes), the
EROS1 data set potentially contains interesting events.

As a final remark, one should also consider that
the systematic survey of thousands of extra-galactic stars at high frequency
($\sim 0.1\,\mathrm{Hz}$) as proposed in this paper may give rise to surprises.
\section{Conclusions}
The minimal condition to observe
an extra-galactic star scintillating
through a Galactic molecular cloud is the
existence of a non vanishing second order derivative of
the optical path in the transverse plane.
To fulfil this condition in the optical domain, stochastic
column density fluctuations of 
$10^{19}\,\mathrm{molecules/cm^2}$
on the $10^3-10^4\,\mathrm{km}$ length scale should take place,
that produce optical path fluctuations of a fraction of a
wavelength.
It follows from existing models of molecular extended objects that
this corresponds to column density relative fluctuations of
a few $ppm$ per $10^3-10^4\,\mathrm{km}$ of transverse distance,
of the same order than the average gradient.

Observations show that structuration of matter
is present at all scales, and certainly do not refute the
eventuality of stochastic fluctuations
producing diffractive scintillation.

In this paper, we showed that there is an
observational opportunity which results from
the subtle compromise between the arm-lever of
interference patterns due to hypothetic diffusive
objects in the Milky-Way and the size of the extra-galactic stars.

The hardware and software techniques required for such observations
are currently available. Tests are under way to validate some of
the concepts described here.
If no technical obstacle arises, there is a true opportunity
to investigate such effects.
\begin{acknowledgements}
The preparation of this paper has benefited from fruitful
discussions and interactions with:
R. Ansari, M.A. Bizouard, E. Falgarone, J. Haissinski, P. Hello,
N. Palanque-Delabrouille, S. Rahvar and P. Schwemling.
Special thanks to J. Haissinski whose remarks allowed to
considerably improve the manuscript.
\end{acknowledgements}

\end{document}